# Variable Neighborhood Search Based Algorithm for University Course Timetabling Problem

**Velin Kralev, Radoslava Kraleva**
*South-West University "Neofit Rilski", Blagoevgrad, Bulgaria*

***Abstract***: *In this paper a variable neighborhood search approach as a method for solving combinatory optimization problems is presented. A variable neighborhood search based algorithm for solving the problem concerning the university course timetable design has been developed. This algorithm is used to solve the real problem regarding the university course timetable design. It is compared with other algorithms that are tested on the same sets of input data. The object and the methodology of study are presented. The main objectives of the experiment are formulated. The conditions for conducting the experiment are specified. The results are analyzed and appropriate conclusions are made. The future trends of work in this field are presented.*

## 1. INTRODUCTION

In solving the combinational optimization problems usually several different algorithms are used that are tested on the same input data. For each specific problem and even for each different set of input data it is necessary to experiment and determine which algorithm and under what conditions (input parameters) would yield the best results for an acceptable time.

For solving the university course timetabling problem in [3] two algorithms are proposed, respectively genetic (GA) and mimetic (MA). The algorithms are tested on real data, obtaining good results. GA quickly finds an acceptable solution evaluation. In contrast, MA finds better solutions, but requires more computational time. Both algorithms are implemented in a real information system for automatic generation of a university course timetable, which is presented in [5]. Also, there is an available web service (presented in [4]), through which input data sets (used to compile a real university course timetable) can be downloaded. On these sets of input data different algorithms (not necessarily heuristics) can be tested that can be adapted and used to solve the studied problem.

In heuristic approaches it is typical that there is a space for many acceptable solutions [3]. Thus the problem is reduced to finding the optimal solution for a given criterion for optimality or a near optimal solution.

The focus in this paper falls on examining the approach VNS (Variable Neighborhood Search) and the development of an algorithm based on this approach. The algorithm must use the valuation model of university course timetable proposed in [2]. This will allow the algorithm to be integrated into the existing information system presented in [5] and to be compared with the already implemented algorithms in it (in the case of GA and MA).

## 2. THE VARIABLE NEIGHBORHOOD SEARCH APPROACH

Variable Neighborhood Search (VNS) is metaheuristic approach which aims to reduce the local minimum and greatly increase the local maximum by changing the adjacent elements when searching for solutions. This approach is often used in cluster analysis, theory of schedules, artificial intelligence and others.

VNS approach was first introduced in [7]. There are various modifications such as: Variable Neighborhood Descent (VND), Reduced VNS (RVNS), Basic VNS (BVNS), Skewed VNS (SVNS), General VNS (GVNS), VN Decomposition Search (VNDS), Parallel VNS (PVNS), Primal Dual VNS (P-D VNS), Reactive VNS, Backward-Forward VNS etc. [6].

The main idea of the VNS approach [1] is described below.

Consider the following optimization problem:

(1) $$\min f(x)$$

(2) $$\min x \in X, X \subseteq S$$

where $f(x)$ is the real objective function which should be minimized. The $X$ is a set of feasible solutions, $x$ is an acceptable solution and $S$ is a space of solutions.

If $S$ is a finite set, which is usually large, then combinatorial optimization problem is defined.

The solution $x^* = X$ is optimal if

(3) $$f(x^*) \leq f(x), \forall x \in X$$

For the problem to be solved it is necessary to find an optimal solution $x^*$. If such an optimal solution doesn't exist then $X = \varnothing$.

When looking for solution, often it is allowed a tolerance, i.e. calculations stop if an acceptable solution $x^*$ is found such that

(4) $$f(x^*) \leq f(x) + \varepsilon, \forall x \in X$$

or

(5) $$\frac{f(x^*) - f(x)}{f(x^*)} \leq \varepsilon, \forall x \in X$$

where $\varepsilon$ is an acceptable tolerance value.

Sometimes an approximate solution is quicly found by using heuristics approaches, meaning that the resulting solution $x_h$ is satisfactory:

(6) $$\frac{f(x_h) - f(x)}{f(x_h)} \leq \varepsilon, \forall x \in X$$

for some $\varepsilon$, which is often large. But the optimality of the solution is not verified.

On the other hand, heuristic methods, which aim to avoid excessive computation time, often face another problem - finding a local optimum.

A local optimum of (1) – (2) is:

(7) $$f(x_L) \leq f(x), \forall x \in N(x_L) \cap X$$

where $N(x_L)$ are neighborhood of $x_L$.

## 3. VARIABLE NEIGHBORHOOD SEARCH BASED ALGORITHM FOR UNIVERSITY COURSE TIMETABLING PROBLEM

The idea of the algorithm is as follows:

**Step 1:** Distribute $N$ events $(n_1, n_2, ..., n_N)$ in $k$ neighborhood structures $(NS_1, NS_2, ..., NS_k)$, so that $k \geq 2$ and $k \leq N/2$. Consider the case in which there are at least two structures formed and in each structure there are at least two events, i.e. $|NS_i| \geq 2, i = 1, 2, ..., k$.

**Step 2:** Select the first structure, $i = 1$ (i.e. $NS_1$).

**Step 3:** Using a method based on local search to find a solution. If a solution is found, it is evaluated and its cost is stored in a temporary list. Furthermore, and the number (index) of the event is recorded, which is the first position in the structure. If the method based on local search does not find a solution, go to step 4.

**Step 4:** Rearrange the events in the current structure, shifted left (or right) with one position (see Fig. 1).

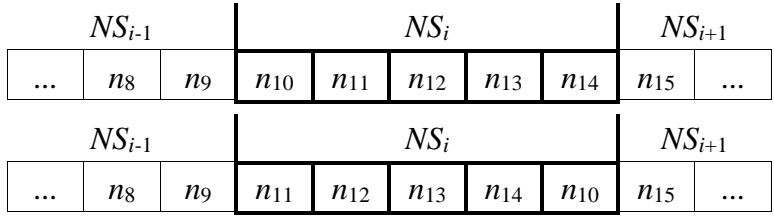

Fig. 1: Shift events in i-th structure with one position left.

If any of the events has visited the first position in the structure (i.e., has turned a full circle), go to step 5, otherwise go to step 3.

**Step 5:** Rearrange the events in the current structure, with such shifts to the left or right that the first position is occupied by the event which had a solution with the best cost (according to the criteria used for optimality, it is solution in which the cost is lower). The list for temporary storage of costs and indices events is cleared.

**Step 6:** Choose the next $NS_{i+1}$ structure, i.e. $(i = i+1)$, and go to Step 3. If there are no more structures, i.e. $i > k$, then End.

If applied, the method based on the local search on the last order of events in the structures, and then the resulting cost will be the best for the so-formed structures. It is possible that this cost is achieved at an early stage of the algorithm, and then improvement is not found.

The pseudo-code of this algorithm is presented in Fig. 2.

```
begin
 N events are divided into k structures NS[][] //k ≥ 2, k ≤ N/2
 for each structure NS[i][] do //i := 1 to k
   for each event in current structure NS[i][j] do //j := 1 to |NSi|
     // beginning of the method based on local search
     for each event n do // n := 1 to N
       if event is fixed then go to next event //n := n + 1
       for each timeslot TS[t] do //t := 1 to T
         try to put the event n in the timeslot TS[t]
         if it is not possible go to next timeslot TS[t+1] // t := t + 1
         if the event is placed, calculate the cost of the solution:
         Cost(CurrentSalution), and store it: SaveCost(BestSalution),
         after that go to the next timeslot //t := t + 1
       end //for each timeslot
       put current event n in this timeslot TS[t], wherein the
       intermediate cost of the solution (BestSalutionCost) was the best,
       then go to the next event //n := n + 1
     end //for each event n
     // end of the method based on local search
     save the cost of finding solution (Solution) to a temporary list L
     L.cost.Add(Cost(Solution))
```

```
      save in L the index j of the event n, which is the first position in
      the current structure NS[i], L.index.Add(NSi[1].j))
      then go to the next event in NS[i][j] // j := j + 1
   end // for each event in current structure
   find the best cost among the saved in L
   best cost := min(L.cost);
   best index = L[best cost].index;
   Rearrange the events in the current structure NS[i], so that the first
   position is the event, in which cost is the best
   for j := 1 to Length(NSi) do
   begin
      if (j = best index) then break else NSi[j].index := NSi[j].index - 1;
      if NSi[j].index = 0 then NSi[j].index = NSi[Length(NSi)].index;
   end
   clear list for temporary storage of costs and indices Clear(L)
   then go to the next structure //NS[i+1]
end //for each structure
with this arrangement of events in structures, generate solution
by the method based on local search
end //of the algorithm
```

Fig. 2. Pseudo-code for generating a solution based on VNS approach.

In performing the method based on local search for each event $n$ each time interval $t$ is checked. The found solution is evaluated by function Cost (Solution), which has linear complexity $\Theta(N)$ depending on the number of events $N$. Since the method based on the local search iterated over all events $N$ and placing each of them calls the function Cost (Solution), then the complexity of the entire process suggests that it requires $N.N$ actions i.e. complexity of the method based on local search is quadratic $\Theta(N^2)$.

In calculating the complexity of the entire algorithm one should take into account the fact that in the outer loop $k$ number of times (as the number of formed structures) is iterated, and each iteration of this loop a nested loop is executed through which all elements in the current structure $|NS_i|$ are visited. That is after all the iterations of the outermost loop (for $k$) all events $N$ will be passed exactly once (each event is assigned to exactly one of these $k$ structures). At each step of the nested loops iterating through the events distributed in structures will call the method based on local search. It has a quadratic complexity as mentioned above. Therefore $N.N^2$ operations, i.e. $N^3$ are carried out. Then, operations are performed by storing values (costs of the solutions found and indices of events) that have $\Theta(1)$ complex. At the last stage, a search of the best cost (and index event) for each of the structures formed with subsequent rearrangement events in the respective structures is performed. These processes have the linear complexity $\Theta(Length(NS_1))$ depending on the number of elements in each structure. These operations do not affect significantly the execution time of the algorithm. Therefore, the complexity of the whole algorithm is cubic, i.e. $\Theta(N^3)$, dependent on the number of events $N$.

## 4. EXPERIMENTAL RESULTS

Three experiments with the following objectives were made:
1. To verify the efficiency of the prototype.
2. To determine the influence of the number of the formed neighborhood structures on the quality of the solution on different sets of input data.
3. To make a comparative analysis between the proposed algorithm and other algorithms used to solve the discussed problem. To compare the performance quality of the generated solutions and execution time of these algorithms. Different algorithms to be tested on the same sets of input data.

### *4.1. Improve the existing prototype*

For the purpose of the experiment the prototype used for analysis of GA and MA described in [3] was improved. The algorithm, which was discussed in the previous section was implemented in the prototype and can be used for automated compilation of university course timetables. Using a prototype the planned experiments were made. From the results, the conclusions were made.

### *4.2. Conditions for the experiment*

Experiments were conducted on a PC with 32 bit operating system Windows 7 Professional (Service Pack 1) and the following hardware configuration: Processor: Intel(R) Core(TM) 2 Duo CPU T7500 @ 2.20GHz 2.20GHz; RAM memory: 2.00 GB.

### *4.3. Methodology of the experiment*

For the purposes of the experiment, 3 sets of input data were used, which are presented in [3]. Input data set "N18" with 18 events, 52 students (divided into 4 groups), 10 lecturers and 10 auditoriums (corresponding to 1 course). Input data set "N90" with 90 events, 175 students (divided in 14 groups), 29 lecturers and 18 auditoriums (corresponding to 1 subject). Input data set "N130" with 130 events, 274 students (divided in 21 groups), 37 lecturers and 22 auditoriums (set of 2 subjects with common events).

## 4.4. Results from the experiments conducted

Tab. 1 shows the parameters of GA and MA. The time of execution for a start of a reproduction is also shown.

Tab. 1: Parameters of GA and MA.

| Parameters of GA and MA. | N18 | N90 | N130 |
|---|---|---|---|
| Number of individuals in the population | 32 | 32 | 32 |
| Number of individuals crossing | 16 | 16 | 8 |
| Number of parental pairs | 8 | 8 | 4 |
| Number of received descendants | 8 | 8 | 4 |
| 20% of descendants mutate | 3 | 3 | 2 |
| Number of reproductions (iterations) | 50 | 50 | 50 |
| Number of solutions to 1 iteration | 43 | 43 | 38 |
| Number of solutions for 1 start | 2150 | 2150 | 1900 |
| Execution time of the MA | 0,31 sec. | 11,25 сек. | 27,68 sec. |
| Execution time of the GA | 0,03 sec. | 0,14 сек. | 0,22 sec. |

After conducting the experiments with GA and MA the results shown in Tab. 2 were obtained.

Tab. 2: Results of the GA and MA..

| GA-N18 | | MA-N18 | | GA-N90 | | MA-N90 | | GA-N130 | | MA-N130 | |
|---|---|---|---|---|---|---|---|---|---|---|---|
| # | Cost | # | Cost | # | Cost | # | Cost | # | Cost | # | Cost |
| 1 | 24,85 | 1 | 1,19 | 1 | 89,92 | 1 | 13,49 | 1 | 151,53 | 1 | 19,72 |
| 2 | 25,51 | 2 | 1,16 | 2 | 89,00 | 2 | 15,16 | 2 | 117,75 | 2 | 21,16 |
| 3 | 24,90 | 3 | 1,27 | 3 | 88,36 | 3 | 13,21 | 3 | 114,90 | 3 | 22,77 |
| 4 | 24,09 | 4 | 1,36 | 4 | 91,34 | 4 | 12,44 | 4 | 117,36 | **4** | **18,47** |
| 5 | 26,05 | 5 | 1,25 | 5 | 89,93 | 5 | 15,03 | 5 | 116,72 | 5 | 20,85 |
| 6 | 25,00 | 6 | 1,19 | 6 | 87,59 | **6** | **11,65** | 6 | 111,95 | 6 | 20,32 |
| 7 | 23,44 | **7** | **1,15** | 7 | 87,71 | 7 | 12,28 | 7 | 113,01 | 7 | 19,12 |
| 8 | 23,37 | 8 | 1,18 | **8** | **86,17** | 8 | 12,29 | 8 | 113,30 | 8 | 20,64 |
| **9** | **22,89** | 9 | 1,21 | 9 | 87,97 | 9 | 11,93 | **9** | **111,38** | 9 | 21,43 |
| 10 | 24,34 | 10 | 1,24 | 10 | 89,33 | 10 | 13,52 | 10 | 117,15 | 10 | 19,02 |

The results shown are the performance of GA and MA in 10 starts for each of the input data sets. The input data set N18 (of 21500 total solutions for both algorithms) shows that the MA best solution is obtained from the seventh starting and has a cost of 1.15. For the GA the best solution is obtained from the ninth starting, which has a cost of 111.38. For input data N90 (also of total solutions 21500 for each of the two algorithms) shows that the MA best solution is obtained from the sixth starting, which has a cost of 11.65. For the GA the best solution is obtained at the eighth start, respectively having cost 86.17. The third input data set N130 (from a total number of 19000 solutions for each of the algorithms) shows that the MA best solution is obtained from the fourth starting, which has a cost of 18.47.

For the GA the best solution is obtained from the ninth starting, which has a cost of 111.38.

After conducting experiments with VNS-based algorithm, tested on input data set N18, the results of it are presented in Tab. 3. The execution time for $k = 2$ is 0.03 sec.

Tab. 3: Results of VNS-based algorithm for input data set N18 (a) $k = 2$, b) $k = 3$, c) k=6).

| k = 2, |NSk| = 9 | | | |
|---|---|---|---|
| NS1 | Cost | NS2 | Cost |
| 1 | 2,85 | 10 | 2,16 |
| 2 | 2,87 | 11 | 1,85 |
| 3 | 2,85 | **12** | **1,85** |
| 4 | 2,85 | 13 | 5,91 |
| 5 | 2,85 | 14 | 3,74 |
| 6 | 3,61 | 15 | 5,16 |
| 7 | 3,61 | 16 | 5,14 |
| 8 | 2,18 | 17 | 2,90 |
| **9** | **2,16** | 18 | 2,85 |

a)

| k = 3, |NSk| = 6 | | | | | |
|---|---|---|---|---|---|
| NS1 | Cost | NS2 | Cost | NS3 | Cost |
| 1 | 2,85 | 7 | 2,84 | **13** | **2,16** |
| 2 | 2,86 | 8 | 2,18 | 14 | 3,58 |
| 3 | 4,30 | 9 | 2,17 | 15 | 3,63 |
| 4 | 3,56 | 10 | 2,17 | 16 | 2,87 |
| 5 | 2,84 | 11 | 2,16 | 17 | 2,87 |
| **6** | **2,84** | **12** | **2,16** | 18 | 2,91 |

b)

| k = 6, |NSk| = 3 | | | | | | | | | | | |
|---|---|---|---|---|---|---|---|---|---|---|---|
| NS1 | Cost | NS2 | Cost | NS3 | Cost | NS4 | Cost | NS5 | Cost | NS6 | Cost |
| 1 | 2,85 | 4 | 2,84 | **7** | **2,84** | 10 | 2,85 | 13 | 2,84 | 16 | 2,84 |
| 2 | 2,85 | 5 | 2,84 | 8 | 4,27 | 11 | 2,84 | 14 | 2,84 | 17 | 2,84 |
| **3** | **2,85** | **6** | **2,84** | 9 | 4,27 | **12** | **2,84** | **15** | **2,84** | **18** | **2,84** |

c)

From Tab. 3 it is seen that for input data set N18, the events are first divided into two structures. The best solution is found for the rearrangement of the events in the first structure when the first position in the structure was an event with index 9. The found solution has cost of 2.16. Then the algorithm is continued searching by rearranging the events in the second structure, the best solution was found when the first position in the second structure has event with index 12. In this arrangement the found solution has cost of 1.85. This is the best solution found so it formed neighborhood structures of events.

In the distribution of events in 3 structures the following results were obtained: at NS1 algorithm has found a solution with the best cost 2.84 when the first position in this structure was event with index 6. Note that this cost is obtained when in the first position in the structure was an event with index 5. In the second structure NS2, the algorithm finds the best solution with cost 2.16 when the first position in this structure was events with indexes 11 and 12 respectively. In rearranging the events of the third structure NS3 there is not an improvement.

In the distribution of events in the 6 structures, the best solution found has a cost of 2.84. It is obtained by rearranging the events in the second structure - NS2. Then no improvement is achieved.

For the input data set N90 were obtained results presented in Tab. 4. Execution time for $k = 2$ is 0.47 sec.

. Tab. 4: Results of VNS-based algorithm for input data set N90.

| k | |NSk| | NSi | # - Index of the event, which is the first position in the structure. C – Solution cost. | | | | | | | | | |
|---|---|---|---|---|---|---|---|---|---|---|---|---|
| 2 | 45 | 1..2 | # | 12 | **58** | | | | | | | | |
| | | | C | 15,37 | **14,75** | | | | | | | | |
| 3 | 30 | 1..3 | # | 12 | 41 | **67** | | | | | | | |
| | | | C | 15,37 | 14,64 | **14,64** | | | | | | | |
| 5 | 18 | 1..5 | # | 12 | 19 | 44 | 72 | **75** | | | | | |
| | | | C | 14,30 | 14,30 | 12,53 | 12,53 | **12,39** | | | | | |
| 6 | 15 | 1..6 | # | 12 | 30 | 33 | 60 | 67 | **77** | | | | |
| | | | C | 12,78 | 12,78 | 12,78 | 12,25 | 12,25 | **12,18** | | | | |
| 9 | 10 | 1..9 | # | 4 | 12 | 25 | 36 | 44 | 52 | 67 | 71 | **88** | |
| | | | C | 15,52 | 14,62 | 14,37 | 13,82 | 13,27 | 12,96 | 12,96 | 12,96 | **12,68** | |
| 10 | 9 | 1..10 | # | 4 | 11 | 19 | 28 | 45 | 46 | 55 | 70 | 70 | **85** |
| | | | C | 16,57 | 15,49 | 15,49 | 15,49 | 15,49 | 15,49 | 15,49 | 14,53 | 13,03 | **13,03** |
| 15 | 6 | 1..10 | # | 2 | 12 | 17 | 19 | 25 | 33 | 39 | 48 | 50 | 58 |
| | | | C | 17,41 | 16,59 | 14,99 | 14,99 | 14,99 | 14,99 | 14,69 | 14,41 | 14,41 | 14,41 |
| | | 11..15 | # | 65 | 70 | 78 | 80 | **85** | | | | | |
| | | | C | 14,17 | 13,31 | 13,31 | 13,31 | **13,31** | | | | | |
| 18 | 5 | 1..10 | # | 3 | 9 | 15 | 20 | 21 | 26 | 33 | 36 | 45 | 50 |
| | | | C | 18,95 | 16,19 | 16,19 | 16,19 | 16,19 | 16,19 | 16,19 | 16,19 | 16,19 | 16,19 |
| | | 11..18 | # | 52 | 60 | 65 | 67 | 75 | 76 | 85 | **89** | | |
| | | | C | 16,19 | 16,19 | 16,19 | 16,19 | 16,19 | 16,19 | 16,19 | **16,19** | | |

Tab. 4 shows that for an input data set N90, the events were divided into 2, 3, 5, 6, 9, 10, 15 and 18 structures. The best found solution is obtained when the events were divided into 6 structures (with 15 events in each structure). In rearranging the events in the last structure NS6, when the first position was an event with index 77, the generated solution has cost of 12.18. Note that the distribution of events in 5 and 9 structures has led to solutions close in cost to the best found, respectively, with costs: at 5 structures: 12,39 and at 9 structures: 12.68.

For the input data set N130 the results presented in Tab. 5 were obtained. Execution time for $k = 2$ is 0.89 sec.

Tab. 5: Results of VNS-based algorithm for input data set N130.

| k | \|NS\| | NSi | # - Index of the event, which is the first position in the structure. C – Solution cost. | | | | | | | | | |
|---|---|---|---|---|---|---|---|---|---|---|---|---|
| 2 | 65 | 1..2 | # | 64 | **67** | | | | | | | |
| | | | C | 21,01 | **21,01** | | | | | | | |
| 5 | 26 | 1..5 | # | 17 | 49 | 71 | 100 | **127** | | | | |
| | | | C | 22,11 | 20,62 | 18,59 | 18,05 | **17,74** | | | | |
| 10 | 13 | 1..10 | # | 12 | 14 | 27 | 45 | 57 | 75 | 91 | 95 | 112 | **130** |
| | | | C | 19,80 | 19,80 | 19,80 | 19,26 | 18,51 | 18,40 | 18,40 | 17,47 | 17,21 | **17,21** |
| 13 | 10 | 1..10 | # | 6 | 20 | 28 | 39 | 43 | 51 | 67 | 73 | 83 | 95 |
| | | | C | 22,44 | 22,44 | 20,61 | 18,80 | 18,80 | 18,80 | 18,80 | 17,87 | 17,80 | 16,86 |
| | | 11..13 | # | 101 | 112 | **130** | | | | | | |
| | | | C | 16,86 | 16,86 | **16,86** | | | | | | |
| 26 | 5 | 1..10 | # | 3 | 9 | 14 | 20 | 21 | 26 | 34 | 40 | 44 | 48 |
| | | | C | 23,99 | 21,63 | 20,81 | 20,81 | 20,81 | 20,81 | 20,81 | 20,64 | 20,64 | 20,64 |
| | | 11..20 | # | 55 | 57 | 64 | 67 | 75 | 76 | 85 | 90 | 94 | 97 |
| | | | C | 20,14 | 19,92 | 19,77 | 19,77 | 18,84 | 18,84 | 18,84 | 18,84 | 18,84 | 18,84 |
| | | 21..26 | # | 105 | 110 | 115 | 120 | 125 | **127** | | | |
| | | | C | 18,84 | 18,84 | 18,84 | 18,84 | 18,84 | **18,84** | | | |

Tab. 5 shows that for an input data set N130, the events were divided into 2, 5, 10, 13 and 26 structures. The best found solution is obtained when the events were divided into 10 structures (with 13 events in each structure). In rearranging the events in structure NS10, when the first position was an event with index 95, the generated solution has cost of 16.86. This cost has not been improved in the next structures. Note that the distribution of events in 5 and 10 structures has led to solutions close in cost to the best found, respectively, with costs: at 5 structures: 17,74 and at 10 structures: 17.21.

### *4.5. Conclusions from the experiments*

After conducting the experiments the following conclusions were made:
1. The developed algorithm can be used in solving the university course timetabling problem.
2. The number of formed neighborhood structures of events, affect the quality of the solution for all sets of input data. Fig. 3a and 3b show graphs of the input data set N90 and N130.

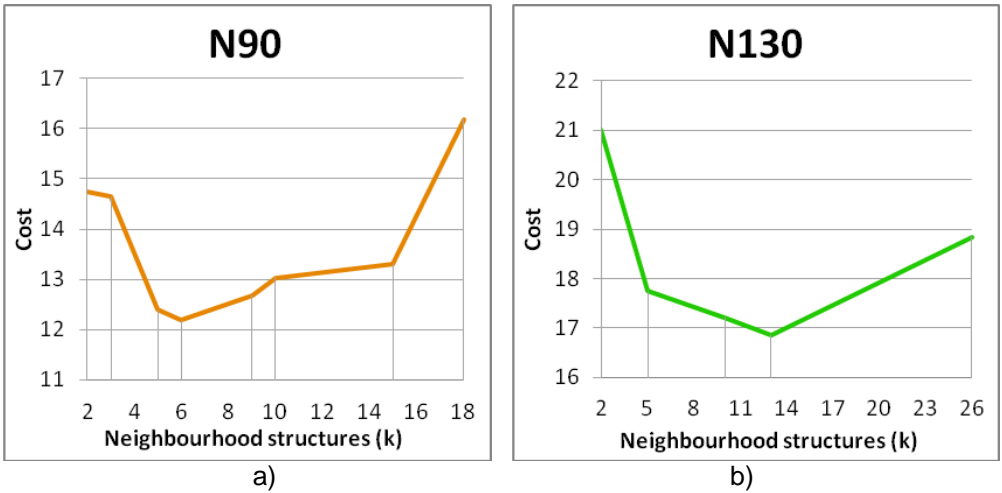

Fig. 3: Behavior of VNS-based algorithm for different number of formed neighborhood structures of events, for the input data set N90 - a) and for N130 - b).

The graph shows that a small number of structures with more events and a large number of structures with few events get the worst results among all the results obtained. From the obtained experimental results, it can be concluded that for N90 best solutions are obtained when each of the formed structures is between 8,1% and 16,2% of the total number of events (in this case 90 events). For N130 best solutions are obtained when each of the formed structures is between 13% and 16.9% of the total number of events (in this case 130).

3. Following the experiment the results obtained in [3] are confirmed. It can be seen that MA gives much better results than GA for all sets of input data, but it requires more CPU time. VNS-based algorithm generates solutions that are much better than those obtained from DA and are commensurate with the solutions obtained from MA. For input data set N130 a solution is found which has a cost of 16.86, which is better than the best solution found by the MA, respectively, with a cost of 18.47.

By increasing the number of events execution time of GA and VNS-based algorithm increases slightly. In contrast, MA requires much more processor time to perform in large input data sets (see Fig. 4).

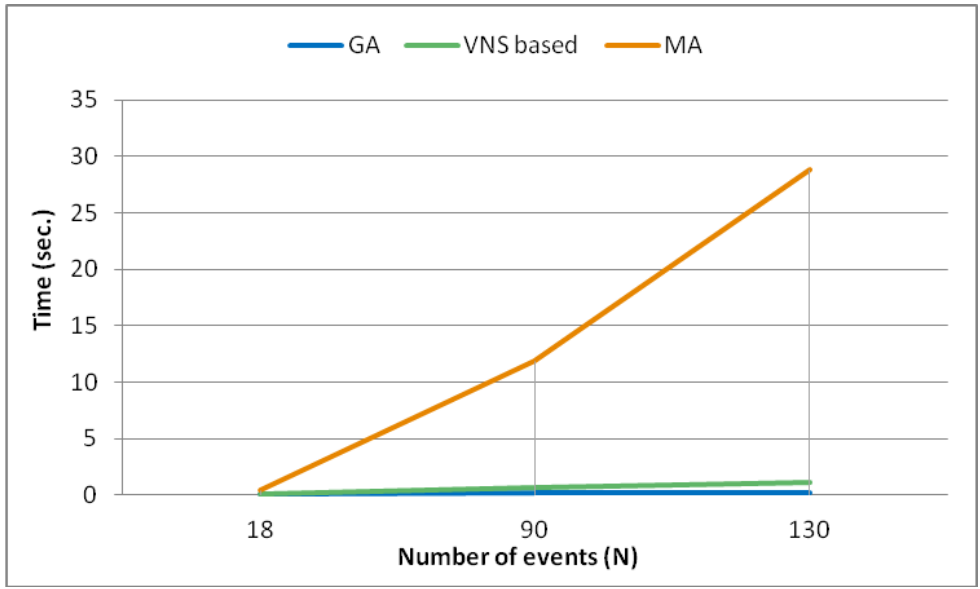

Fig. 4: Execution time of algorithms with increasing number of events.

## 5. CONCLUSION

In this article VNS based algorithm is presented. It was successfully used to solve the university course timetabling problem. The methodology and the object of the study are presented. Also, the objectives of the planned experiments are formulated. The proposed algorithm is compared with two other algorithms - GA and MA. All algorithms are tested on the same input data sets.

In conclusion, we note that the developed VNS based algorithm, can be used successfully in solving the university course timetabling problem as well as in small input data sets and large ones. By conducting experiments it is seen that for small input data sets, MA gives better results in a reasonable time. By increasing the number of events, however, the execution time of MA is much greater than the time for VNS-based algorithm. Furthermore, for the same input data sets it finds comparable solutions, and in some cases better than those found by the MA.

As future trends of employment seeking ways to optimize the algorithm in terms of execution time may be noted. It is also necessary to do a large-scale research to determine the optimum ratio between the number of the formed neighborhood structures, and the number of events in each structure.

# 6. REFERENCES

[1] Hansen, P., Mladenović, N. (2003) A Tutorial on Variable Neighborhood Search, Les Cahiers du GERAD.

[2] Kralev, V. (2009) A Model for the University Course Timetable Problem. International Journal "Information Technologies & Knowledge" Vol.3, pp 276-289.

[3] Kralev, V. (2009) A Genetic and Memetic Algorithm for Solving the University Course Timetable Problem. International Journal "Information Theories & Applications" Vol.16, pp 291-299.

[4] Kralev, V., Kraleva, R. (2011) Web Service Based System for Generating Input Data Sets. Proceedings of the Fourth International Scientific Conference – FMNS2011, Vol. 2, pp 49-56.

[5] Kralev, V., Kraleva, R., Siniagina, N. (2009) An Integrated System for University Course Timetabling. Proceedings of the Third International Scientific Conference – FMNS2009, Vol. 1, pp 99-105.

[6] Mladenović, N. (2012) Recent advances in Variable neighborhood search, GOW-2012, Brazil (http://gow12.dca.ufrn.br/files/gow12-Mladenovic.pdf).

[7] Mladenović, N., Hansen, P. (1997) Variable neighborhood search, Computers and Operations Research, Vol. 24, Issue 11, Elsevier Science, pp. 1097–1100.